\documentclass[aps,amssymb,pra,twocolumn,showpacs,showkeys,floatfix]{revtex4}

\usepackage{epsf}
\usepackage{amsmath}
\usepackage{graphicx}
\begin{document}
         
\title {Gamma radioactivity of anomalous wells}

\author{Boris I. Ivlev}

\affiliation
{Instituto de F\'{\i}sica, Universidad Aut\'onoma de San Luis Potos\'{\i},
San Luis Potos\'{\i}, 78000 Mexico\\}

\begin{abstract}

Gamma emission of nuclear energy scale ($\sim 3MeV$), caused by electron transitions in anomalous wells, is predicted to occur in acoustic experiments with solids. The anomalous well for 
electrons is formed by a local reduction of electromagnetic zero point energy in a vicinity of a nucleus which can be a lattice site of a solid \cite{IVLEV3}. The well width is $\sim 10^{-11}cm$ 
and the well depth is $\sim 3MeV$. An energy spectrum in anomalous wells is continuous and non-decaying. Unusual experimental results, on unexpected emission from lead of  $\sim 1keV$ x-rays under 
acoustic pulses, are likely explained by formation of anomalous wells \cite{LOSK}. The experimentally observed $keV$ quanta are naturally supplemented by $MeV$ emission to be revealed. This 
conclusion is drawn on the basis of an exact solution within a model generic with quantum electrodynamics. An energy of emitted quanta (x-rays and gamma) comes from a reduction of 
electromagnetic zero point energy (energy from ``nothing''). 

\end{abstract} \vskip 1.0cm



\maketitle

\section{Introduction}
\label{intr}
Discrete energy levels of an electron in a potential well become of a finite width under the interaction with photons. This provides finite 
lifetimes of levels with respect to photon emission. Photons are emitted until the electron reaches the ground state level. This level has zero 
width. It is slightly shifted due to the electron-photon interaction (the Lamb shift) \cite{LANDAU3,WELT,MIGDAL,KOLOM}. Discrete levels can be of different lifetimes. When the certain level is 
long-living this results in possibility of laser generation. Broadening of the energy level is characterized by the imaginary part of its energy. The level widths in various atoms are known 
\cite{KRAUSE}. The corresponding energy dissipation provides friction motion of the particle. This problem is generic with one in Ref.~\cite{LEGG} studying dissipative quantum mechanics. See also
\cite{LAR1,SCHMID,MELN,CHAKR,HANG1,LEGG1,KOR,IVLEV1,HANG2,KAGAN,LAR2,WEISS,MAG1,MAG2,MAG3,MAG4}.

As shown in Ref.~\cite{IVLEV2} on the basis of an exact analytical solution, besides such dissipative motion there is another scenario of particle-photon states. That new regime is not dissipative.
The joint particle-photon state is stationary and elastic deformations accompany the particle. This reminds polaronic state in a solid when the electron is ``dressed'' by phonons \cite{KITT}. 
Our polaronic states in the well are continuously distributed in energy which does not have an imaginary part. This means that the polaronic states at any energy are non-decaying. 

Analogous electron states (continuous in energy and with no dissipation) occur in anomalous wells created by local reduction of electromagnetic zero point energy \cite{IVLEV3}. The width of this 
well is $\sim 10^{-11}cm$ and the well depth is $\sim 3MeV$. The well is formed during time interval $\hbar/3MeV\sim 10^{-22}s$. Wells, produced by a local reduction of electromagnetic zero point 
energy, are also formed in the Casimir effect \cite{CAS,LANDAU3}. But in this case they are shallow and widely extended in space. 

Since states in the well are non-decaying, there is a non-trivial question: (i) will the particle go down in energy due to nonstationary effects after a well formation or (ii) it will be ``frozen'' 
at finite energies. As shown in this paper, the answer corresponds to (i). The particle reaches a low energy in the well. The related non-conservation of energy is due to a work of the source 
which provides the well formation. This source is electromagnetic zero point energy which varies in space providing the anomalous well. Therefore a transition of electron down in energy is 
assisted by gamma emission with the energy $\sim 3MeV$ (energy from ``nothing''). This estimate follows from the depth of the anomalous well.

Unusual experimental results \cite{LOSK} are likely explained by formation of anomalous wells for electrons. In those experiments strong acoustic pulses, acted on lead, resulted in unexpected 
x-rays of $\sim 1keV$ energy. Lead nuclei moved fast, under acoustic perturbations, so their de Broglie wave length became short corresponding to the size of anomalous wells stimulating their 
formation in vicinity of lead nuclei. It was a table-top experiment. 

In a usual lead atom a mean binding energy per electron is a few $keV$ \cite{LANDAU4,LOSK}. After the fast (within the time $\hbar/3MeV\sim 10^{-22}s$) formation of the anomalous well a 
rearrangement of the usual electrons is not equally fast. They transfer into upper levels of an anomalous well during longer time $\hbar/1keV\sim 10^{-18}s$ resulting in the observed x-rays of 
the $keV$ energy. After that the electrons go to lower levels in the deep anomalous well emitting $\sim 3MeV$ gamma quanta. So the both ($keV$ and $MeV$) photon emissions are from the same 
track. The $keV$ part was already observed but the $MeV$ part is to be revealed.

It is amazing that the emission of gamma quanta of a nuclear energy can occur in table top acoustic experiments with lead. A mechanism of this emission is not nuclear but electronic.

One can put a question about energy production by that gamma radiation (energy from ``nothing''). It was briefly discussed already in Refs.~\cite{IVLEV3,LOSK} when a possibility of 
gamma emission was not completely clear. Now gamma emission is shown to be real. In experiments \cite{LOSK} solely $10^{-2}g$ part of the lead sample was effectively involved into the 
gamma production. If to specially rearrange a lead matter, one can reach, say, $100g$ part of the lead to be effective. This increases the gamma yield up to four orders of magnitude making 
promising energy production in those acoustic processes. That issue can be significant for practical applications.

\section{PARTICLE WITH DISSIPATION}
\label{ela}
Suppose the total system to consist of a particle of mass $m$ in the harmonic potential $m\Omega^2x^2/2$ and an elastic string placed along the $z$ axis and described by transverse displacements
$v(z,t)$. The classical Lagrangian of this system 
\begin{equation}
\label{1}
{\cal L}_0=\frac{m}{2}\left(\dot x^2-\Omega^2x^2\right)+\frac{\rho}{2}\int dz\bigg[\dot v^2-c^2\left(\frac{\partial v}{\partial z}\right)^2\bigg]
\end{equation}
depends on a mass density $\rho$ of the string and contains the elastic string term proportional to $(\partial v/\partial z)^2$. 

It is convenient to consider the string of the finite length $L$. In the finale results the limit $L\rightarrow\infty$ is taken. According to this, one can use the Fourier series
\begin{equation}
\label{3}
v(z,t)=\frac{1}{L}\sum_{n}v_n(t)\exp\left(\frac{2\pi in}{L}z\right),
\end{equation}
where all $n$ are involved and $v_{-n}=v^{*}_{n}$. In the representation (\ref{3}) the Lagrangian (\ref{1}) takes the form
\begin{equation}
\label{4}
{\cal L}_0=\frac{m}{2}\left(\dot{x}^2-\Omega^2x^2\right)+\frac{\rho}{2L}\sum_n\left(\dot{v}_{n}\dot{v}_{-n}-\omega^{2}_{n}v_{n}v_{-n}\right),
\end{equation}
where $\omega_n=2\pi |n|s/L$.

In the system, described by (\ref{4}), the particle and the elastic string are independent. But our goal is to introduce a particle-string interaction just to produce a particle dissipation. An
adequate way to do that is to use the Caldeira-Leggett approach \cite{LEGG} accounting a particle-string interaction by the Lagrangian 
\begin{equation}
\label{5}
{\cal L}={\cal L}_0-x\sum_nc_nv_n-x^2\sum_n\frac{Lc_nc_{-n}}{2\rho\omega^{2}_{n}},
\end{equation}
where $c_{-n}=c^{*}_{n}$. According to \cite{LEGG}, the last term in (\ref{5}) is added due to renormalization just to get a physical result. 

An $n$ dependence of $c_n$ is connected to dissipative properties of the system \cite{LEGG}. To specify $c_n$ we consider a classical limit in Sec.~\ref{class}. 
\subsection{Classical limit} 
\label{class}
The classic equations of motion, following from the Lagrangian (\ref{5}), are
\begin{eqnarray}
\nonumber
&&m\ddot x+m\tilde\Omega^2x=-\sum_nc_nv_n\\
&&\frac{\rho}{L}\left(\ddot v_n+\omega^{2}_{n}v_n\right)=-xc_{-n},
\label{6}
\end{eqnarray}
where $\tilde\Omega^2=\Omega^2+\sum_nLc_nc_{-n}/(\rho\omega^{2}_{n})$. It follows from (\ref{6})
\begin{equation}
\label{7}
v_n(t)=-\frac{Lc_{-n}}{\rho\omega^{2}_{n}}\int^{t}_{-\infty}dt_1x(t_1)\sin\omega_n(t-t_1).
\end{equation}
Substituting this result in (\ref{6}), it is easy to show that
\begin{eqnarray}
\nonumber
&&m\ddot x+m\Omega^2x=\\
&&-\frac{1}{\pi}\int^{t}_{-\infty}dt_1\dot x(t_1)\int^{\infty}_{-\infty}d\omega_n\eta_1(\omega_n)\cos\omega_n(t_1-t),
\label{8}
\end{eqnarray}
where
\begin{equation}
\label{9}
\eta_1(\omega_n)=\frac{L^2c_nc_{-n}}{2\rho \omega^{2}_{n}}. 
\end{equation}
The Fourier component of Eq.~(\ref{8}) reads
\begin{eqnarray}
\label{10}
&&m(\Omega^2-\omega^2)x_{\omega}-\\
&&i\omega x_{\omega}\int^{\infty}_{-\infty}\frac{d\omega_n}{\pi}\,\eta_1(\omega_n)\int^{0}_{-\infty}d\tau e^{-i\omega\tau}\cos\tau\omega_n=0.
\nonumber
\end{eqnarray}
Under the $\tau$-integration one should treat $\omega$ as $\omega+i\delta$ which results in
\begin{eqnarray}
\label{11}
&&m(\Omega^2-\omega^2)x_{\omega}-\omega x_{\omega}\int^{\infty}_{-\infty}\frac{d\omega_n}{2\pi}\,\eta_1(\omega_n)\\
&&\times\left(\frac{1}{\omega_n-\omega-i\delta}-\frac{1}{\omega_n+\omega+i\delta}\right)=0.
\nonumber
\end{eqnarray}

Now the classical dynamic equation takes the form
\begin{equation}
\label{12}
m(\Omega^2-\omega^2)x_{\omega}+i\omega\eta(\omega)x_{\omega}=0,
\end{equation}
where the total friction coefficient is $\eta(\omega)=\eta_1(\omega)+i\eta_2(\omega)$ and, according to dispersion relations \cite{LANDAU2},
\begin{equation}
\label{13}
\eta_2(\omega)=\frac{\omega}{\pi}\int^{\infty}_{-\infty}d\omega_n\frac{\eta_1(\omega_n)-\eta_1(\omega)}{\omega^2-\omega^{2}_{n}}.
\end{equation}
So the real part of the friction coefficient $\eta_1(\omega)$ determines the the entire complex $\eta(\omega)$.

Zeros in the complex $\omega$-plane of the expression (\ref{12}) are equivalent to poles of the generalized susceptibility \cite{LANDAU2}. Those poles cannot be in the upper half plane of the
complex $\omega$ due to the causality principle. This restricts possible functional $\omega$-dependence of $\eta(\omega)$. In classical electrodynamics 
\begin{equation}
\label{14}
\eta(\omega)=\frac{2e^2}{3c^3}\omega^2,\hspace{0.5cm}\omega\rightarrow 0
\end{equation}
corresponds to Bremsstrahlung \cite{LANDAU1}. According to that, one can choose a model leading to the correct electrodynamic result (\ref{14}). In this model
\begin{equation}
\label{16}
\eta(\omega)=\frac{\eta_0\omega^2}{(\omega_0-i\omega)^2},\hspace{0.5cm}\eta_0=\frac{2e^2\omega^{2}_{0}}{3c^3}
\end{equation}
and
\begin{equation}
\label{17}
\eta_1(\omega)=\eta_0\frac{\omega^2(\omega^{2}_{0}-\omega^2)}{(\omega^{2}_{0}+\omega^2)^2},\hspace{0.5cm}\eta_2(\omega)=\eta_0\frac{2\omega_0\omega^3}{(\omega^{2}_{0}+\omega^2)^2}.
\end{equation}
It is easy to check that with the choice (\ref{16}) all zeros of (\ref{12}) (poles of the generalized susceptibility) are in the lower half plane of complex $\omega$. 

The frequency $\omega_0$ remains a free parameter. A comparison to quantum electrodynamics says that anyway $\omega_0$ is no more than $mc^2$.

\section{REDUCTION TO INDEPENDENT OSCILLATORS}
\label{indep}
It is convenient to make a transformation in the classical Lagrangian (\ref{5})
\begin{equation}
\label{18}
v_n=\xi_n-\frac{Lxc_{-n}}{\rho\omega^{2}_{n}}.
\end{equation}
Now the Lagrangian (\ref{5}) reads
\begin{eqnarray}
\label{19}
&&{\cal L}=\frac{m}{2}\left(1+\sum_{n}\beta_n\beta_{-n}\right)\dot x^2-\frac{m\Omega^2}{2}x^2+\\
&&\frac{\rho}{2L}\sum_n\left(\dot\xi_n\dot\xi_{-n}-\omega^{2}_{n}\xi_n\xi_{-n}\right)-\dot x\sqrt{\frac{m\rho}{L}}\sum_n\beta_n\dot\xi_n,
\nonumber
\end{eqnarray}
where
\begin{equation}
\label{20}
\beta_n=\frac{c_n}{\omega^{2}_{n}}\sqrt{\frac{L}{m\rho}}\,.
\end{equation}
We remind that the summation in (\ref{19}) is extended on $-\infty<n<\infty$. 

The Lagrangian (\ref{19}) consists of two quadratic forms of velocities and coordinates. There are no cross terms. Therefore one can make a transformation of variables $\{\xi_n\}, x$ into the
certain new coordinates $\{\eta_p\}$ 
\begin{equation}
\label{21}
\xi_n=\sum_pu_{np}\eta_p,\hspace{0.5cm}x=\sqrt{\frac{\rho}{mL}}\sum_p\tilde u_p\eta_p,
\end{equation}
when the two quadratic forms are diagonal, that is
\begin{equation}
\label{22}
{\cal L}=\frac{m}{2}\sum_p\left(\dot\eta^{2}_{p}-\omega^{2}_{p}\eta^{2}_{p}\right),
\end{equation}
where $\omega_p$ is a function to be determined. After a substitution of the new variables into the Lagrangian (\ref{19}) we arrive to
\begin{equation}
\label{23}
{\cal L}=\frac{\rho}{2L}\sum_{p,q}\left(A_{pq}\dot\eta_p\dot\eta_q-B_{pq}\eta_p\eta_q\right),
\end{equation}
where
\begin{eqnarray}
\nonumber
&&A_{pq}=\left(1+\sum_n\beta_n\beta_{-n}\right)\tilde u_p\tilde u_q+\sum_nu_{np}u_{-nq}\\
&&-\sum_n\beta_n\left(\tilde u_pu_{nq}+\tilde u_{q}u_{np}\right)
\label{24}
\end{eqnarray}
and
\begin{equation}
\label{25}
B_{pq}=\Omega^2\tilde u_p\tilde u_q+\sum_n\omega^{2}_{n}u_{np}u_{-nq}\,.
\end{equation}
The diagonal form (\ref{23}) corresponds to the conditions
\begin{equation}
\label{26}
A_{pq}=\frac{mL}{\rho}\delta_{pq},\hspace{0.5cm}B_{pq}=\frac{mL}{\rho}\,\omega^{2}_{p}\delta_{pq}.
\end{equation}
\subsection{Spectrum of independent oscillators}
\label{spectr}
The relation $\omega^{2}_{q}A_{pq}=B_{pq}$, which follows from (\ref{26}), has the form
\begin{eqnarray}
\label{27}
&&\left[\omega^{2}_{q}\left(1+\sum_n\beta_n\beta_{-n}\right)-\Omega^2\right]\tilde u_p\tilde u_q+\\
&&\sum_n\left(\omega^{2}_{q}-\omega^{2}_{n}\right)u_{np}u_{-nq}=\omega^{2}_{q}\sum_n\beta_n\left(\tilde u_{q}u_{np}+\tilde u_pu_{nq}\right).
\nonumber
\end{eqnarray}
Equalizing the coefficient at $u_{np}$ to zero we obtain
\begin{equation}
\label{28}
\left(\omega^{2}_{q}-\omega^{2}_{n}\right)u_{-nq}=\omega^{2}_{q}\sum_n\beta_n\tilde u_q.
\end{equation}
Analogously, equalizing in (\ref{27}) the coefficient at $\tilde u_p$ to zero, one obtains 
\begin{equation}
\label{29}
\left[\omega^{2}_{q}\left(1+\sum_n\beta_n\beta_{-n}\right)-\Omega^2\right]\tilde u_q=\omega^{2}_q{\sum_n}\beta_nu_{nq}.
\end{equation}
The relations, following from Eq.~(\ref{28}),
\begin{equation}
\label{30}
u_{nq}=\frac{\omega^{2}_{q}\beta_{-n}}{\omega^{2}_{q}-\omega^{2}_{n}}\,\tilde u_q
\end{equation}
should be substituted into (\ref{29}). The result is 
\begin{eqnarray}
\nonumber
&&\frac{\Omega^2}{\omega^{2}_{q}}-1+\frac{2c}{mL}\sum_{n}\frac{\eta_1(\omega_n)-\eta_1(\omega_q)}{\omega^{2}_{q}-\omega^{2}_{n}}\\
&&=\frac{2c}{mL}\eta_1(\omega_q)\sum_n\frac{1}{\omega^{2}_{n}-\omega^{2}_{q}}
\label{31}
\end{eqnarray}
Here we account for the expression of $\beta_n\beta_{-n}$ through $\eta_1(\omega_n)$ using (\ref{9}) and (\ref{20}). 

In the right-hand side of Eq.~(\ref{31}) one has to use the relation
\begin{equation}
\label{32}
\sum_n\frac{1}{n^2-a^2}=-\frac{\pi}{a}\cot\pi a.
\end{equation}
In the sum in the left-hand side of (\ref{31}) the summation can be substituted by the integration $\sum_n\rightarrow L/(2\pi c)\int d\omega_n$ ($-\infty<\omega_n<\infty$). With the expression 
(\ref{13}) Eq.~(\ref{31}) takes the final form
\begin{equation}
\label{33}
m\left(\omega^{2}_{q}-\Omega^2\right)-\omega_q\eta_2(\omega_q)=\omega_q\eta_1(\omega_q)\cot\frac{L\omega_q}{2c}\,.
\end{equation}
Eq.~(\ref{33}) determines the function $\omega_q$ in the Lagrangian (\ref{22}). It has the form
\begin{equation}
\label{34}
\omega_q\rightarrow\omega_q+\frac{2c}{L}\left[\frac{\pi}{2}+\arctan\frac{m(\Omega^2-\omega^{2}_{q})+\omega_q\eta_2}{\omega_q\eta_1}\right],
\end{equation}
where in the right-hand side $\omega_q=2\pi c|n|/L$. 
\subsection{Calculation of $u_{np}$ and $\tilde u_p$}
\label{param}
As follows from (\ref{25}) and (\ref{30}), 
\begin{equation}
\label{35}
\frac{B_{pq}}{\tilde u_p\tilde u_q}=\Omega^2+\omega^{2}_{p}\omega^{2}_{q}\sum_n\frac{\omega^{2}_{n}\beta_n\beta_{-n}}{(\omega^{2}_{n}-\omega^{2}_{p})(\omega^{2}_{n}-\omega^{2}_{q})}\,.
\end{equation}
The right-hand side of (\ref{35}) at $p\neq q$ 
\begin{equation}
\label{36}
\Omega^2+\frac{\omega^{2}_{p}\omega^{2}_{q}}{\omega^{2}_{p}-\omega^{2}_{q}}\sum_n\omega^{2}_{n}\beta_n\beta_{-n}
\left(\frac{1}{\omega^{2}_{n}-\omega^{2}_{p}}-\frac{1}{\omega^{2}_{n}-\omega^{2}_{q}}\right)
\end{equation}
is zero due to the relation (\ref{31}). Therefore the Lagrangian in new variables has the diagonal form (\ref{22}). The parameter $B_{pp}$ is given by (\ref{26}) and this allows to calculate 
$\tilde u_p$ from Eq.~(\ref{35}). By means of Eqs.~(\ref{9}) and (\ref{20}), $\tilde u_p$ is defined by 
\begin{equation}
\label{37}
\frac{mL\omega^{2}_{p}}{\rho\tilde u^{2}_{p}}=\Omega^2+\frac{2c\omega^{4}_{p}}{mL}\sum_n\frac{\eta_1(\omega_n)}{(\omega^{2}_{n}-\omega^{2}_{p})^2}.
\end{equation}
We do not perform detailed calculations in Eq.~(\ref{37}) because $\tilde u_p$ is not required below. 

\section{NONSTATIONARY POTENTIAL WELL}
\label{nonst}
We used above the Lagrangian formalism to reduce the system to a set of independent oscillators. This also can be done within the Hamiltonian formalism. In this case the Lagrangian (\ref{19}) 
generates the Hamiltonian consisting of a quadratic form of momenta and a quadratic form of coordinates. The old momenta are transformed through new ones in a linear way and the old 
coordinates are transformed through new ones also in a linear way. The transformation is canonical and the coefficients in the linear forms depend on $\Omega$. 

The same canonical transformation can be done when $\Omega$ depends on time because the coefficients in the linear forms remain the same with $\Omega(t)$ as a parameter. Indeed, in that canonical
transformation those coefficients are not differentiated by time. 

Instead of the static potential $m\Omega^2x^2/2$ in the Lagrangian (\ref{4}) we choose now 
\begin{equation}
\label{38}
V(x,t)=\frac{m\Omega^2(t)}{2}x^2-\varepsilon(t).
\end{equation}
The corresponding Schr\"{o}dinger equation for the wave function $\psi(\{\eta_q\},t)$ takes the form
\begin{equation}
\label{39}
i\hbar\frac{\partial\psi}{\partial t}=\sum_q\left[-\frac{\hbar^2}{2m}\frac{\partial^2}{\partial\eta^{2}_{q}}+\frac{m\omega^{2}_{q}(t)}{2}\eta^{2}_{q}\right]\psi-\varepsilon(t)\psi,
\end{equation}
where time dependence of $\omega_q(t)$ is determined by (\ref{34}) with a nonstationary $\Omega(t)$. 

The wave function has the form
\begin{equation}
\label{40}
\psi\left(\{\eta_q\},t\right)=\exp\left[\frac{i}{\hbar}\int^{t}_{-\infty}\varepsilon(t_1)dt_1\right]\prod_q\psi_q(\eta_q,t),
\end{equation}
where $\psi_q(\eta_q,t)$ satisfies the equation
\begin{equation}
\label{41}
i\hbar\frac{\partial\psi_q}{\partial t}=-\frac{\hbar^2}{2m}\frac{\partial^2\psi_q}{\partial\eta^{2}_{q}}+\frac{m\omega^{2}_{q}(t)}{2}\eta^{2}_{q}\psi_q\,.
\end{equation}
\subsection{Wave function}
\label{solut}
Below we consider parameters in the potential (\ref{38}) $\Omega(t)=\Omega\theta(t)$ and $\varepsilon(t)=\varepsilon\theta(t)$ where the parameters in right-hand sides are constants. This choice
corresponds to an ``instant'' well creation. In this case $\omega_q(t)=\omega^{+}_{q}\theta(t)+\omega^{-}_{q}\theta(-t)$, where $\omega^{+}_{q}$ is determined by Eq.~(\ref{34}). The frequency
$\omega^{-}_{q}$ is defined by Eq.~(\ref{34}), where one has to put $\Omega=0$.

To construct a solution of Eq.~(\ref{41}) one should know a solution of the classical equation
\begin{equation}
\label{42}
\ddot z_q+\omega^{2}_{q}(t)z_q=0.
\end{equation}
The solution of (\ref{42}) is
\begin{equation}
\label{43}
z_q(t)=\sqrt{\frac{\hbar}{m\omega^{-}_{q}}}
\begin{cases}
\exp(it\omega^{-}_{q}),& t<0\\
\exp(it\omega^{+}_{q})+i\left(\frac{\omega^{-}_{q}}{\omega^{+}_{q}}-1\right)\sin t\omega^{+}_{q},& 0<t
\end{cases}
\end{equation}
If to represent $z_q=r_q\exp(i\gamma_q)$ then 
\begin{equation}
\label{44}
r_q(t)=\sqrt{\frac{\hbar}{m\omega^{-}_{q}}}
\begin{cases}
1,& t<0\\
1+(\omega^{-}_{q}/\omega^{+}_{q}-1)\sin^2 t\omega^{+}_{q},& 0<t
\end{cases}
\end{equation}
and
\begin{equation}
\label{45}
\gamma_q(t)=
\begin{cases}
t\omega^{-}_{q},& t<0\\
t\omega^{+}_{q}+(\omega^{-}_{q}/\omega^{+}_{q}-1)\sin t\omega^{+}_{q}\cos t\omega^{+}_{q},& 0<t
\end{cases}
\end{equation}
In Eqs.~(\ref{43}) - (\ref{45}) the parts $(\omega^{+}_{q}-\omega^{-}_{q})$ are small since they are proportional to $1/L$. A solution of the Schr\"{o}dinger equation (\ref{41}) has the 
form \cite{PER} 
\begin{equation}
\label{46}
\psi_q(\eta_q,t)=\frac{1}{\sqrt{r_q}}\exp\left(\frac{im\dot r_q}{2\hbar r_q}\eta^{2}_{q}\right)
\chi_q\left(\frac{\eta_q}{r_q}\sqrt{\frac{\hbar}{m\omega^{-}_{q}}}\,,\,\frac{\gamma_q}{\omega^{-}_{q}}\right),
\end{equation}
where the function $\chi_q(v,\tau)$ satisfies the equation
\begin{equation}
\label{47}
i\hbar\frac{\partial\chi(v,\tau)}{\partial\tau}=-\frac{\hbar^2}{2m}\frac{\partial^2\chi}{\partial\tau^2}+\frac{m(\omega^{-}_{q})^2}{2}v^2\chi\,.
\end{equation}
An advantage of this solution is that it allows to match the region before and after appearance of the well. The solutions we need is
\begin{equation}
\label{48}
\chi_q=\frac{1}{\sqrt{r_q\sqrt{\pi}}}\exp\left(-\frac{\eta^{2}_{q}}{2r^{2}_{q}}-\frac{i\gamma_q}{2}\right).
\end{equation}
\subsection{Energy of the state}
\label{energ}
\begin{figure}
\includegraphics[width=5.0cm]{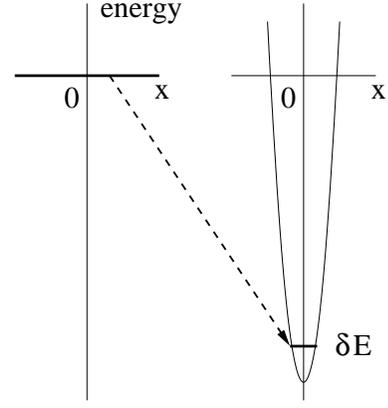}
\caption{\label{fig1}Energy additional to zero point one $\sum_q\hbar\omega^{-}_{q}/2$. (Left) A particle interacts with photons in absence of an external potential. (Right) The same particle in 
the potential $m\Omega^2x^2/2-\varepsilon$ which is suddenly formed. The transition occurs to the energy $\delta E$.}
\end{figure}

At $t<0$ the expression (\ref{48}) goes over into the usual ground state wave function with the energy $E_q=\hbar\omega^{-}_{q}/2$. But at $0<t$ it becomes non-trivial and the energy $E_q$ 
acquires an additional part. Generally
\begin{equation}
\label{49}
E_q=i\hbar\int\psi^{*}_{q}\frac{\partial\psi_q}{\partial t}d\eta_q.
\end{equation}
To perform calculations in (\ref{49}) one has to use Eqs.~(\ref{46}) and (\ref{48}) with the definitions (\ref{44}) and (\ref{45}). One should not differentiate $r_q$ in (\ref{48}) since this
provides zero total contribution. Note that in our case $\langle\eta^{2}_{q}\rangle=r^{2}_{q}/2$. So in a calculation of the time derivative in (\ref{49}) one has to differentiate in the wave 
function solely $\dot r_q$ in (\ref{46}) and $\gamma_q$ in (\ref{48}). It is easy to show that the result is
\begin{equation}
\label{50}
E_q=\frac{\hbar}{2}\dot\gamma_q-\frac{\ddot r_q}{4}\sqrt{\frac{m\hbar}{\omega^{-}_{q}}}=\frac{\hbar\omega^{-}_{q}}{2}\theta(-t)+\frac{\hbar\omega^{+}_{q}}{2}\theta(t).
\end{equation}
Note that despite $\dot\gamma_q$ and $\ddot r_q$ in (\ref{50}) are functions of $t$ the result is time independent excepting the jumps in time. The total energy can be written in the form
\begin{equation}
\label{51}
E=\sum_q\frac{\hbar\omega^{-}_{q}}{2}+\delta E\theta(t),
\end{equation}
where the first part is simply a total energy of zero point oscillations at $t<0$. The second term in (\ref{51}) 
\begin{equation}
\label{52}
\delta E=\sum_q\frac{\hbar}{2}\left(\omega^{+}_{q}-\omega^{-}_{q}\right)-\varepsilon
\end{equation}
corresponds to a formation of the potential well $m\Omega^2x^2/2-\varepsilon$ at $t=0$. One can substitute in (\ref{52}) $\sum_q\rightarrow L/(2\pi c)\int d\omega_q$ ($0<\omega_q<\infty$) resulting 
in
\begin{eqnarray}
\nonumber
&&\delta E=\hbar\int^{\infty}_{0}\frac{d\omega}{2\pi}\bigg[\arctan\frac{m\omega-\eta_2(\omega)}{\eta_1(\omega)}\\
&&-\arctan\frac{m(\omega^2-\Omega^2)-\omega\eta_2(\omega)}{\omega\eta_1(\omega)}\bigg]-\varepsilon.
\label{53}
\end{eqnarray}
The integral in (\ref{53}) can be easily evaluated since $\eta_1$ and $\eta_2$ are small. At $\omega<\Omega$ the both parts of the integrand in (\ref{53}) are $\pi/2$ but at $\Omega<\omega$ the
integrand is small. Therefore
\begin{equation}
\label{54}
\delta E\simeq\frac{\hbar\Omega}{2}-\varepsilon.
\end{equation}
Corrections to this result are on the order of $e^2/\hbar c$. In Fig.~\ref{fig1} the transition to the energy $\delta E$ is shown. The energy spectrum in the well in Fig.~\ref{fig1} is 
continuous and non-decaying. $\delta E$ is the mean energy to where the transition occurs. A non-conservation of system energy is due to a work of the external source which provides the well 
formation.
\subsection{Particle without dissipation}
\label{nodiss}
It is instructive to know what happens to transitions after an instant formation of the well in the case when the particle is not connected to photons. In this case the particle is described by 
the Schr\"{o}dinger equation
\begin{equation}
\label{55}
i\hbar\frac{\partial\psi}{\partial t}=-\frac{\hbar^2}{2m}\frac{\partial^2\psi}{\partial x^2}+\left[\frac{m\Omega^{2}(t)}{2}x^2-\varepsilon\theta(t)\right]\psi,
\end{equation}
where $\Omega(t)=\Omega^{-}\theta(-t)+\Omega^{+}\theta(t)$. Analogously to Sec.~\ref{solut}, a solution of Eq.~(\ref{55}) has the form
\begin{equation}
\label{56}
\psi(x,t)=\frac{1}{\sqrt{r\sqrt{\pi}}}\exp\left(\frac{im\dot r}{2\hbar r}x^2-\frac{x^2}{2r^2}-\frac{i\gamma}{2}+\frac{i\varepsilon t}{\hbar}\right),
\end{equation}
where $r(t)$ and $\gamma(t)$ are given by Eqs.~(\ref{44}) and (\ref{45}) if to make formal substitutions $\omega^{-}_{q}\rightarrow\Omega^{-}$ and $\omega^{+}_{q}\rightarrow\Omega^{+}$. The energy 
of the state
\begin{equation}
\label{57}
E=i\hbar\int\psi^*\frac{\partial\psi}{\partial t}dx=\frac{\hbar}{2}\dot\gamma+\frac{m}{4}\left(\dot r^2-r\ddot r\right)-\varepsilon\theta(t)
\end{equation}
turns into
\begin{equation}
\label{58}
E=
\begin{cases}
\frac{\hbar\Omega^-}{2},&t<0\\
\frac{\hbar\Omega^+}{2}-\varepsilon+\frac{\hbar(\Omega^+-\Omega^-)^2}{2\Omega^-}\cos^2t\Omega^+,&0<t
\end{cases}
\end{equation}
In contrast to the case of particle-photon interaction (\ref{50}), the particle from the ground state $\hbar\Omega^-/2$ after the instant well formation does not go to the new ground state 
$\hbar\Omega^+/2-\varepsilon$. Instead, it oscillates over a finite interval of the energy spectrum. This follows from the last (interference) term in (\ref{58}) at $0<t$. When initially the 
particle was free ($\Omega^-\rightarrow 0$) that interval becomes infinite. The nonstationary energy at $0<t$ results from a superposition of states with various eigenenergies.

\section{DISCUSSIONS}
\label{disc}
A particle in a well has discrete energy levels. Under interaction with photons those levels acquire a finite width resulting in loose of a particle energy which finally reaches a ground state 
value. In contrast to such dissipative state another type of states in a well is possible which reminds a polaronic state in solids \cite{KITT}. In this case bound particle-photon states are of 
continuous energy spectrum. They are also stationary that is with zero imaginary parts of energy (no photon emission despite the particle is not in a ground state). Existence of such states in a 
well is shown on a basis of an exact analytical solution \cite{IVLEV2}. 

One can qualitatively explain why photons are not emitted by polaronic states. Emission of waves would result also in oscillations of the particle which is coupled to photons. This increases the 
particle kinetic energy preventing it to lose its total energy and therefore resulting in non-decaying states. These general arguments do not depend on a type of the potential where the particle 
moves. For this reason, the polaronic states may exist not in harmonic potentials only. 

Suppose that initially there is no well and the particle interacts solely with photons and the total system is in a ground state. This state of the particle-photon system is exact one. Then 
suppose that at $t=0$ a well is instantly formed. Since states in the well are non-decaying there is a non-trivial question: (i) will the particle go down in energy due to nonstationary effects 
after a well formation or (ii) it will be ``frozen'' at finite energies. As shown in this paper, the answer corresponds to (i). The particle reaches the low energy in the well which is approximately 
the ground state energy if to switch off the interaction with photons. That non-conservation of energy is due to a work of the source which provides the well formation.

The model, considered in the paper, corresponds to Caldeira-Leggett approach \cite{LEGG} with particle-photon interaction similar to one in quantum electrodynamics. This model enables to extend
the obtained results to a significant problem related to electrons in anomalous wells \cite{IVLEV3}. The well is formed by a local reduction of electromagnetic zero point energy in a vicinity of a
heavy nucleus. The well width is $\sim 10^{-11}cm$ and the well depth is $\sim 3MeV$ \cite{IVLEV3}. An energy spectrum in anomalous wells is continuous and non-decaying. 

Wells, produced by a local reduction of electromagnetic zero point energy, are also formed in the Casimir effect \cite{CAS,LANDAU3}. But in this case they are shallow and widely extended in space.

Unusual experimental results \cite{LOSK} are likely explained by formation of anomalous wells. In those experiments strong acoustic pulses, acted on lead, resulted in unexpected x-rays of the 
energy $\sim 1keV$. Heavy lead nuclei moved fast, under acoustic perturbations, so their de Broglie wave length became short corresponding to the size of anomalous wells stimulating their 
formation in vicinity of lead nuclei. It was a table-top experiment. 

An anomalous well is formed during the time interval $\hbar/3MeV\sim 10^{-22}s$. This is practically instant compared to typical electron time in solids. So this is close to the situation described
in the paper and therefore one can generally use the conclusion drawn. A cause of anomalous well formation is not an artificial source, considered in the paper, but an electromagnetic system providing
a local redistribution of its zero point energy. Therefore, the particle transition to the lower level, as in Fig.~\ref{fig1}, is accompanied by a gamma emission of the energy $\sim 3MeV$. This 
estimate follows from the depth of the anomalous well. 

We emphasize that the mechanism of the gamma emission is not due to a usual photon-induced broadening of discrete energy levels. There is no such broadening in our system. gamma quanta are
emitted due to nonstationary processes of well formation. After the well formation is finished the particle takes its stationary position close to well bottom. In contrast, without coupling to
photons the particle oscillates over the energy spectrum after well formation. 

In a usual lead atom a mean binding energy per electron is a few $keV$ \cite{LANDAU4,LOSK}. After the fast (within the time $\hbar/3MeV\sim 10^{-22}s$) formation of the anomalous well a 
rearrangement of the usual electrons is not equally fast. They transfer into upper levels of an anomalous well during longer time $\hbar/1keV\sim 10^{-18}s$ resulting in the observed x-rays of 
the $keV$ energy. After that the electrons go to lower levels in the deep anomalous well emitting $\sim 3MeV$ gamma quanta. So the both ($keV$ and $MeV$) photon emissions are from the same 
track. The $keV$ part was already observed but the $MeV$ part is to be revealed.

It is amazing that the emission of gamma quanta of a nuclear energy can occur in table top acoustic experiments with lead. A mechanism of this emission is not nuclear but electronic.

One can put a question about energy production by that gamma radiation (energy from ``nothing''). It was briefly discussed already in Refs.~\cite{IVLEV3,LOSK} when a possibility of 
gamma emission was not completely clear. Now gamma emission is shown to be real. In experiments \cite{LOSK} solely $10^{-2}g$ part of the lead sample was effectively involved into the 
gamma production. If to specially rearrange a lead matter, one can reach, say, $100g$ part of the lead to be effective. This increases the gamma yield up to four orders of magnitude making 
promising energy production in those acoustic processes. That issue can be significant for practical applications.

\section{CONCLUSIONS}
The anomalous well for electrons is formed by a local reduction of electromagnetic zero point energy in a vicinity of a nucleus. The well width is $\sim 10^{-11}cm$ and the well depth is 
$\sim 3MeV$. An energy spectrum in anomalous wells is continuous and non-decaying \cite{IVLEV3}. 

Unusual experimental results, on unexpected emission from lead of $\sim 1keV$ x-rays under acoustic pulses, are likely explained by formation of anomalous wells \cite{LOSK}.

It is shown that gamma emission ($\sim 3MeV$) from anomalous wells is expected. This conclusion is drawn on the basis of an exact solution within a model generic with quantum electrodynamics. 
An energy of emitted quanta (x-rays and gamma) comes from a reduction of an electromagnetic zero point energy (energy from ``nothing''). gamma emission of nuclear energy scale is predicted 
to occur in table top acoustic experiments.

\acknowledgments
I appreciate stimulating discussions with I. Pavlova. My general view on the problem was formed with an influence of A. M. Loske. This work was supported by CONACYT through grant 237439.


\begin{thebibliography}{8}

\bibitem{IVLEV3}

B. I. Ivlev, arXiv:1701.00520

\bibitem{LOSK}

F. Fern\'{a}ndez, A. M. Loske, and B. I. Ivlev, arXiv:1804.00530

\bibitem{LANDAU3}

V. B. Berestetskii, E. M. Lifshitz, and L. P. Pitaevskii {\it Quantum Electrodynamics} (Pergamon, New York, 1980).

\bibitem{WELT}

T. A. Welton, Phys. Rev. {\bf 74}, 1157 (1948).

\bibitem{MIGDAL}

A. B. Migdal, {\it Qualitative Methods in Quantum Theory} (Addison-Wesley, 2000).

\bibitem{KOLOM}

E. B. Kolomeisky, arXiv:1203.1260

\bibitem{KRAUSE}

M. O. Krause and J. H. Oliver, J. Phys. Chem. Ref. Data {\bf 8}, No. 2 (1979).

\bibitem{LEGG}

A. O. Caldeira, and A. J. Leggett, Annals of Physics {\bf 149}, 374 (1983).

\bibitem{LAR1}

A. I. Larkin and Yu. N. Ovchinnikov, Zh. Eksp. Teor. Fiz. Pis'ma, {\bf 37}, 322 (1983) [Sov. Phys. JETP Lett.{ \bf 37}, 382 (1983)].

\bibitem{SCHMID}

A. Schmid, Phys. Rev. Lett. {\bf 51}, 1506 (1983).

\bibitem{MELN}

V. I. Melnikov, Zh. Eksp. Teor. Fiz. {\bf 87}, 663 (1984) [Sov. Phys. JETP { \bf 60}, 380 (1984)].

\bibitem{CHAKR}

L. -D. Chang and S. Chakravarty, Phys. {\bf 29}, 130 (1984).

\bibitem{HANG1}

H. Grabert, U. Weiss, and P. H\"{a}nggi,  Phys. Rev. Lett. {\bf 51}, 1506 (1983).

\bibitem{LEGG1}

A. J. Leggett, S. Chakravarty, A. T. Dorsey, M. P. A. Fisher, A. Garg, and W. Zwerger, Rev. Mod. Phys. {\bf 59}, 1 (1987).

\bibitem{KOR}

S. E. Korshunov, Zh. Eksp. Teor. Fiz. {\bf 92}, 1828 (1987) [Sov. Phys. JETP { \bf 66}, 872 (1987)].

\bibitem{IVLEV1}

B. I. Ivlev, Zh. Eksp. Teor. Fiz. {\bf 94}, 333 (1988) [Sov. Phys. JETP { \bf 68}, 1486 (1988)].

\bibitem{HANG2}

P. H\"{a}nggi, P. Talkner, and M. Borcovec, Rev. Mod. Phys. {\bf 62}, 251 (1990).

\bibitem{KAGAN}

Yu. Kagan and N. V. Prokof'ev, in {\it Quantum Tunneling in Condensed Media}, edited by A. Leggett and Yu. Kagan (North-Holland, Amsterdam, 1992).

\bibitem{LAR2}

A. I. Larkin and Yu. N. Ovchinnikov, in {\it Quantum Tunneling in Condensed Media}, edited by A. Leggett and Yu. Kagan (North-Holland, Amsterdam, 
1992).

\bibitem{WEISS}

U. Weiss, in {\it Series in Modern Condensed Matter Physics}, vol. 2 (World Scientific, Singapore 1993). 

\bibitem{MAG1}

L. Magazz\`{u}, D. Valenti, A. Carollo, and B. Spagnolo, Entropy {\bf 17}, 2341 (2015).

\bibitem{MAG2}

D. Valenti, L. Magazz\`{u}, P. Caldara, and B. Spagnolo, Phys. Rev. B {\bf 91}, 235412 (2015).

\bibitem{MAG3}

L. Magazz\`{u}, D. Valenti, B. Spagnolo, and M. Grifoni, Phys. Rev. E {\bf 92}, 032123 (2015).

\bibitem{MAG4}

L. Magazz\`{u}, A. Carolo, B. Spagnolo, and D. Valenti, J. Stat. Mechanics: Theory and Experiment 054016 (2016).

\bibitem{IVLEV2}

B. I. Ivlev, Can. J. Phys. {\bf 94}, 1253 (2016).

\bibitem{KITT}

C. Kittel, {\it Quantum Theory of Solids} (John Wiley and Sons, New York, 1963).

\bibitem{CAS}

H. B. G. Casimir and D. Polder, Phys. Rev. {\bf 73}, 360 (1948).

\bibitem{LANDAU4}

L. D. Landau and E. M. Lifshitz, {\it Quantum Mechanics} (Pergamon, New York, 1977).

\bibitem{LANDAU2}

L. D. Landau and E. M. Lifshitz, {\it Statistical Physics} (Butterworth-Heinemann, 1980).

\bibitem{LANDAU1}

L. D. Landau and E. M. Lifshitz, {\it The Classical Theory of Fields} (Pergamon, New York, 1975).

\bibitem{PER}

A. M. Perelomov and Ya. B. Zeldovich, {\it Quantum Mechanics (Selected Topics)} (World Scientific Publishing, 1998).


\end{thebibliography}
\end{document}